\documentclass[12pt]{article}
\usepackage{graphicx}

\begin{document}

\baselineskip=14pt plus 1pt minus 1pt

\begin{center}

{\Large\bf Complementary descriptions of shape/phase transitions 
in atomic nuclei} 

\bigskip

{Dennis Bonatsos$^{a}$\footnote{e-mail: bonat@inp.demokritos.gr},
E. A. McCutchan$^{b}$\footnote{e-mail:
elizabeth.ricard-mccutchan@yale.edu},
N. V. Zamfir$^{c}$\footnote{e-mail: zamfir@tandem.nipne.ro}
}
\medskip

{$^{a}$ Institute of Nuclear Physics, N.C.S.R. ``Demokritos'',}

{GR-15310 Aghia Paraskevi, Attiki, Greece}

{$^{b}$ Wright Nuclear Structure Laboratory, Yale University,}

{New Haven, Connecticut 06520-8124, USA}

{$^{c}$ National Institute of Physics and Nuclear Engineering, 
Bucharest-Magurele, Romania} 

\end{center}

\bigskip
\centerline{\bf Abstract} \medskip

Shape/phase transitions in atomic nuclei have first been discovered 
in the framework of the Interacting Boson Approximation (IBA) model. 
Critical point symmetries appropriate for nuclei at the transition points 
have been introduced as special solutions of the Bohr Hamiltonian, 
stirring the introduction of additional new solutions describing wide 
ranges of nuclei. The complementarity of the IBA and geometrical approaches 
will be demonstrated by three examples. First, it will be shown that 
specific special solutions of the Bohr Hamiltonian correspond to the borders
of the critical region of the IBA. Second, it will be demonstrated that
the distinct patterns exhibited in different transitional regions by the 
experimental energy staggering in $\gamma$-bands can be reproduced 
both by the IBA and by special solutions of the Bohr Hamiltonian. 
Third, a first attempt to obtain a IBA SU(3) level scheme from a special 
solution of the Bohr Hamiltonian will be presented.  

\section{Introduction} 

Atomic nuclei are known to exhibit changes of their energy levels and 
electromagnetic transition rates among them when the number of protons 
and/or neutrons is modified, resulting in shape phase transitions from one 
kind of collective behaviour to another. These transitions are not phase 
transitions of the usual thermodynamic type. They are quantum phase 
transitions \cite{IacIJMPA}
(initially called ground state phase transitions \cite{Deans}), occurring 
in Hamiltonians of the type 
$ H = c(H_1 + g H_2)$,
where $c$ is a scale factor, $g$ is the control parameter, and $H_1$, $H_2$ 
describe two different phases of the system. The expectation value of a 
suitably chosen operator, characterizing the state of the system, is used as
the order parameter.

In the framework of the Interacting Boson Approximation (IBA) model \cite{IA}, which describes 
nuclear structure of even--even nuclei within the U(6) symmetry, possessing 
the U(5), SU(3), and O(6) limiting dynamical symmetries, appropriate for 
vibrational, axially deformed, and $\gamma$-unstable nuclei respectively, 
shape phase transitions have been studied 25 years ago \cite{Deans}
using the classical limit of the 
model \cite{GK,GK2,DSI,vanRoos}, 
pointing out that there is (in the usual Ehrenfest classification) 
a second order shape phase transition between U(5) and O(6), a first order 
shape phase transition between U(5) and SU(3), and no shape phase transition 
between O(6) and SU(3). It is instructive to place \cite{IacIJMPA}
these shape phase transitions 
on the symmetry triangle of the IBM \cite{Casten}, at the three corners of 
which the three limiting symmetries of the IBM appear. 

More recently it has been realized \cite{IacE5,IacX5} that the properties 
of nuclei lying at the critical point of a shape phase transition can be 
described by appropriate special solutions of the Bohr Hamiltonian \cite{Bohr},
labelled as critical point symmetries. The E(5) critical point symmetry 
\cite{IacE5} has been found to correspond to the second order critical point 
between U(5) and O(6), while the X(5) critical point symmetry \cite{IacX5}
has been found to correspond to the first order transition between 
U(5) and SU(3).  

The introduction of the critical point symmetries E(5) \cite{IacE5} and X(5) 
\cite{IacX5} has triggered 
much work on special solutions of the Bohr Hamiltonian, corresponding to 
different physical situations \cite{CastenNP,Poenaru}. Several of these solutions 
will be mentioned below, where the complementarity of the IBA and geometrical approaches 
will be demonstrated by three examples, given in Sections 2, 3, and 4. 

\section{Connecting the X(5)-$\beta^2$, X(5)-$\beta^4$, and
X(3) models to the shape/phase transition region of the
Interacting Boson Model} 

As already mentioned in the introduction, 
shape/phase transitions in atomic nuclei 
were first investigated~\cite{Deans} within the Interacting Boson
Approximation (IBA) model \cite{IA} by constructing the classical
limit of the model, using the coherent state formalism
\cite{GK,GK2,DSI,vanRoos}. Using this method it was shown
\cite{Deans,DSI} that the shape/phase transition between
the U(5) [spherical, vibrational] and SU(3) [prolate axially symmetric 
deformed, rotational] limiting symmetries is of first order, while
the transition between the U(5) and O(6) ($\gamma$-unstable)
limiting symmetries is of second order. Furthermore, the region of
phase coexistence within the symmetry triangle \cite{Casten} of
the IBA has been studied \cite{IZC,Zamfir66,IZ} and its borders
have been determined \cite{Werner,Fernandes}, while a similar
structural triangle for the geometric collective model has been
constructed \cite{Zhang}.

Recently, special solutions of the Bohr Hamiltonian, called
critical point symmetries \cite{IacE5,IacX5}, describing nuclei at
the points of shape/phase transitions between different limiting
symmetries, have attracted considerable attention, since
they lead to parameter independent (up to overall scale factors)
predictions which are found to be in good agreement with
experiment \cite{CZ1,ClarkE5,CZ2,ClarkX5}. The X(5) critical point
symmetry \cite{IacX5}, was developed to describe analytically the
structure of nuclei at the critical point of the transition from
vibrational [U(5)] to prolate axially symmetric [SU(3)] shapes.
The solution involves a five-dimensional infinite square well
potential in the $\beta$ collective variable and a harmonic
oscillator potential in the $\gamma$ variable. 

The success of the
X(5) model in describing the properties of some nuclei with
parameter free (except for scale) predictions has led to
considerable interest in such simple models to describe
transitional nuclei. Since its development, numerous extensions
involving either no free parameters or a single free parameter
have been developed. Those approaches which involve a single
parameter include replacing the infinite square well potential
with a sloped well potential~\cite{markslope}, exact decoupling of
the $\beta$ and $\gamma$ degrees of freedom~\cite{mark2005}, and
displacement of the infinite square well potential, or the
confined $\beta$-soft model~\cite{CBS}. Parameter free variants of
the X(5) model include the X(5)-$\beta^2$ and X(5)-$\beta^4$
models \cite{BonX5}, in which the infinite square well potential
is replaced by a $\beta^2$ and a $\beta^4$ potential respectively,
as well as the X(3) model \cite{X3}, in which the $\gamma$ degree
of freedom is frozen to $\gamma=0$, resulting in a
three-dimensional Hamiltonian, in which an infinite square well
potential in $\beta$ is used.

It is certainly of interest to examine the extent to which the
parameter free (up to overall scale factors) predictions of the
various critical point symmetries and related models, built within
the geometric collective model, are related to the shape/phase
transition region of the IBA. It has already been found
\cite{McC7103} that the X(5) predictions cannot be exactly
reproduced by any point in the two-parameter space of the IBA,
while best agreement is obtained for parameters corresponding to a
point close to, but outside the shape/phase transition region of
the IBA.  

In a recent paper \cite{Libby1} the parameter independent (up to overal scale
factors) predictions of the X(5)-$\beta^2$, X(5)-$\beta^4$, and
X(3) models, which are variants of the X(5) critical point
symmetry developed within the framework of the geometric
collective model, are compared to the results of two-parameter
Interacting Boson Approximation (IBA) model calculations, with the
aim of establishing a connection between these two approaches. It
turns out that both X(3) and X(5)-$\beta^2$ lie close to the
U(5)--SU(3) leg of the IBA symmetry triangle and  within the
narrow shape/phase transition region of the IBA. In particular,
X(3) lies close to $\zeta_{\rm crit}$  \cite{Fernandes}, the left border of the
shaded shape/phase transition region of the IBA, corresponding to
IBA total energy curves with two equal minima, while
X(5)-$\beta^2$ lies near the right border of the shape/phase
transition region, $\zeta^{**}$ \cite{Werner}, corresponding to IBA total energy
curves with a single deformed minimum. A set of neighboring
even-even nuclei exhibiting the X(3), X(5)-$\beta^2$, and
X(5)-$\beta^4$ behaviors have been identified
($^{172}$Os-$^{174}$Os-$^{176}$Os) \cite{Libby1}. Additional examples for X(3),
X(5)-$\beta^2$, and X(5)-$\beta^4$ are found in $^{186}$Pt,
$^{146}$Ce, and $^{158}$Er, respectively \cite{Libby1}.  The level of agreement
of these parameter free, geometrical models with these candidate
nuclei is found to be similar to the predictions of the
two-parameter IBA calculations.

It is intriguing that the X(3) model, which corresponds to an
exactly separable $\gamma$-rigid (with $\gamma=0$) solution of
the Bohr collective Hamiltonian, is found to be related to the IBA
results at $\zeta_{\rm crit}$, which corresponds to the critical
case of two degenerate minima in the IBA total energy curve \cite{Fernandes},
approximated by an infinite square well potential in the model. It
is also remarkable that the X(5)-$\beta^2$ model, which uses 
the same approximate separation of variables as the X(5) critical
point symmetry, is found to correspond to the right border
($\zeta^{**}$) of the shape/phase transition region \cite{Werner}, related to
the onset of total energy curves with a single deformed minimum,
comparable in shape with the $\beta^2$ potential used in the model
in the presence of a $L(L+1)/(3\beta^2)$ centrifugal term
\cite{mark2005}.

Comparisons in the same spirit of the parameter independent
predictions of the E(5) critical point symmetry \cite{IacE5} and
related E(5)-$\beta^{2n}$ models \cite{AriasE5,BonE5}, as well as
of the related to triaxial shapes Z(5) \cite{Z5} and Z(4)
\cite{Z4} models, to IBA calculations and possible placement of
these models on the IBA-1 symmetry triangle (or the IBA-2 phase
diagram polyerdon \cite{AriasIBM2,CIPRL,CIAP}) can be illuminating
and should be pursued.

It should be noticed that Ref. \cite{Libby1} has been focused on
boson numbers equal or close to 10, to which many nuclei
correspond. A different but interesting question is to examine if
there is any connection between the X(3), X(5)-$\beta^2$, and/or
X(5)-$\beta^4$ models and the IBA for large boson numbers. This is
particularly interesting especially since it has been established
(initially for $N=1,000$ \cite{AriasE5}, recently corroborated for
$N=10,000$ \cite{AriasE5b}) that the IBA critical point of the
U(5)-O(6) transition for large $N$ corresponds to the
E(5)-$\beta^4$ model, i.e. to the E(5) model employing a $\beta^4$
potential in the place of the infinite well potential
\cite{AriasE5,BonE5}.

\section{Staggering in $\gamma$ bands and the transition
between different symmetries of nuclear structure} 

The importance of the $\gamma$ degree of freedom in nuclei with
static quadrupole deformation has been known for decades starting
with the work of Wilets and Jean~\cite{wilets}, and Davydov,
Filippov, and Chaban~\cite{triax1,triax2}.  Since the bandhead of
the quasi-$\gamma$ band is not fixed in most geometrical
models, we investigate instead, the parameter free predictions of
the spacings within the quasi-$\gamma$ band. For this purpose, we
use the odd--even staggering in gamma bands \cite{stag}
$$
S(J)={ [E(J_\gamma^+) - E((J-1)_\gamma^+)] - [E((J-1)_\gamma^+) -
E((J-2)_\gamma^+) \over E(2_1^+)},$$
\noindent which measures the displacement of the $(J-1)_\gamma^+$
level relative to the average of its neighbors, $J_\gamma^+$ and
$(J-2)_\gamma^+$, normalized to the energy of the first excited
state of the ground state band, $2_1^+$. Since $S(J)$ is a
(discrete) derivative, it is very sensitive to structural changes.
For example, the energy levels of the $\gamma$ band in a
$\gamma$-independent potential \cite{wilets} cluster as $(2_\gamma^+)$, 
$(3_\gamma^+, 4_\gamma^+)$, \dots, opposite to the rigid triaxial 
rotor $(2_\gamma^+, 3_\gamma^+)$,$(4_\gamma^+, 5_\gamma^+)$, \dots
clustering pattern \cite{triax1,triax2}. 

In a recent paper \cite{Libby2}, three categories of transitional regions 
have been considered: 

1) The $\gamma$-soft region between the vibrator and a deformed $\gamma$-soft
structure where the potential is $\gamma$-independent. This corresponds to the 
U(5) to O(6) transition in the language of the Interacting Boson Approximation 
(IBA)\cite{IA}. This is the region containing the critical point symmetry E(5)
\cite{IacE5}, as well as the second-order phase transition in the IBA between 
U(5) and O(6) \cite{Deans,DSI}. The Xe, Ba, and Ce series of isotopes provide good manifestations 
of this region \cite{IA,Libby2}. 
It is seen that $S(J)$ exhibits strong staggering with minima
at even $J$ and maxima at odd $J$. This behavior is reproduced in the geometrical 
framework by the chain of parameter-independent models formed by E(5) \cite{IacE5}, 
in which an infinite well $u(\beta)$ potential is used, and the E(5)-$\beta^{2n}$
$(n=1,2,3,4)$ models \cite{AriasE5,BonE5}, in which $u(\beta)=\beta^{2n}/2$. In all cases the potential
is $\gamma$-independent. In  the IBA it corresponds to $\chi=0$ and increasing $\zeta$
\cite{Zamfir66,Werner}.  

2) The axially $\gamma$-rigid region between the vibrator and the axially symmetric 
rotor, characterized by a harmonic oscillator in $\gamma$ with the minimum in 
$\gamma$ close to zero. This is the U(5) to SU(3) transition region of the IBA \cite{IA}, 
in which a first-order phase transition occurs \cite{Deans,DSI}. This is also the region where 
the critical point symmetry X(5) \cite{IacX5} is found. Several Sm, Gd, Dy, Er, U, and Fm 
isotopes exhibit this behavior \cite{IA,Libby2}. It is seen that $S(J)$ exhibits weak staggering with minima
at even $J$ and maxima at odd $J$. This behavior is reproduced in the geometrical 
framework by the chain of parameter-independent models formed by X(5) \cite{IacX5}, 
in which an infinite well $u(\beta)$ potential is used, and the X(5)-$\beta^{2n}$
$(n=1,2,3,4)$ models \cite{BonX5}, in which $u(\beta)=\beta^{2n}/2$, with a potential of the form 
$u(\beta)+v(\gamma)$ used in all cases, as well as by the exactly separable ES-X(5)
and ES-X(5)-$\beta^2$ models \cite{ESX5}, in which potentials of the form 
$u(\beta)+v(\gamma)/\beta^2$ are used. In all cases $v(\gamma)$ is a steep harmonic oscillator
potential centered at $\gamma=0^\circ$. In the IBA this region  corresponds 
to $\chi=-1.32 $ and increasing $\zeta$ \cite{Zamfir66,Werner}.   

3) The triaxial $\gamma$-rigid region between the vibrator and the rigid triaxial 
rotator \cite{triax1,triax2}, characterized by fixed $\gamma$-values between $0^\circ$ and $30^\circ$,
which has no analog in the framework of IBA-1 \cite{IA}. Few nuclei ($^{112}$Ru, 
$^{170}$Er, $^{192}$Os, $^{192}$Pt, $^{232}$Th) are known to show this behavior \cite{Libby2}, 
in which $S(J)$ exhibits strong staggering with minima
at odd $J$ and maxima at even $J$. This behavior is reproduced in the geometrical 
framework by the chain of parameter-independent models formed by Z(5) \cite{Z5}, 
 Z(5)-$\beta^{2}$, Z(4) \cite{Z4}, and Z(4)-$\beta^2$, which are special solutions 
of the Bohr Hamiltonian for $\gamma=30^\circ$ in five and four dimensions respectively. 

Having discussed the dependence of $S(J)$ on angular momentum, it is interesting 
to focus attention on $S(4)$,
the relative displacement of the $3_\gamma^+$ state relative to
the average of its neighbors, $2_\gamma^+$ and $4_\gamma^+$,
normalized to the energy of the $2_1^+$ state, and discuss its variation 
as a function of changing structure, using as a structure indicator the 
ratio $R_{4/2}=E(4_1^+)/E(2_1^+)$. The following results are found \cite{Libby2}:

i) In the U(5)-O(6) region the IBA predicts $S(4)$ decreasing with increasing 
$R_{4/2}$. The same prediction is made by the geometrical models mentioned in 1), 
completed with O(5)-CBS \cite{O5CBS}, which corresponds to E(5) with an infinite well 
potential displaced from the origin. The Xe and Ba chains of isotopes provide an 
experimental manifestation of this behavior. 

ii) In the O(6)-SU(3) region the IBA predicts $S(4)$ increasing with increasing 
$R_{4/2}$. The same prediction is made by the geometrical models mentioned in 3). 

iii) In the U(5)-SU(3) region the IBA predicts $S(4)$ decreasing with increasing 
$R_{4/2}$ for low values of $R_{4/2}$, then abruptly jumping to a value close to zero,
close to which it remains with further increase of $R_{4/2}$. The geometrical
models mentioned in 2) reproduce the same behavior (slight increase of $S(4)$ 
with increasing $R_{4/2}$) beyond the abrupt change, a situation corroborated 
by experimental evidence from the Nd, Sm, Gd, Dy, and Er isotope chains \cite{Libby2}. 
The abrupt change appears to be related to the first order phase transition
\cite{Deans,DSI} 
occuring in this region. It appears that phase transitional behavior occurs 
close to where $S(4)$ crosses zero.   

\section{Occurence of IBA SU(3) degeneracy in the geometrical model}  

A long standing problem is the derivation from the Bohr
Hamiltonian \cite{Bohr} of a spectrum similar to that of the SU(3) limit of
the Interacting Boson Approximation (IBA) model \cite{IA}. The
main features of the spectrum should be:

a) The energy spacings among the $2^+$, $4^+$, $6^+$, \dots levels
within the ground, $\beta$ and $\gamma$ bands should be identical.

b) Furthermore, the $2^+$, $4^+$, $6^+$, \dots   levels of the
$\beta$ and $\gamma$ bands should be degenerate.

An attempt to solve this problem has been carried out 
in a recent paper \cite{ESD}, where an exactly separable version of the Bohr
Hamiltonian, called ES-D, which uses a potential of the form
$u(\beta)+u(\gamma)/\beta^2$ \cite{wilets}, with a Davidson potential $\beta^2
+\beta_{0}^4/\beta^2$ 
\cite{Dav,Elliott,Bahri} in the place of $u(\beta)$, and a steep harmonic 
oscillator centered at $\gamma=0^\circ$ as 
$u(\gamma)$, is developed. All bands (e.g., ground, $\beta$ and
$\gamma$) in this model are treated on equal footing \cite{IacCam}, depending on
two parameters, the Davidson parameter $\beta_{0}$ and the
stiffness $c$ of the $\gamma$-potential. The model is found \cite{ESD} to be
applicable only to well deformed nuclei (with $R_{4/2}\geq 3.0$)
due to the $\beta^2$ denominator in the $u$($\gamma$) term.
Nevertheless, it reproduces very well the bandheads and energy
spacings within bands of almost all rare earth and actinide
nuclei, with $R_{4/2}\geq 3.0$,  for which available data exists,
as well as most of the $B(E2)$ transition rates. The most glaring
discrepancy concerns $B$($E$2) values for the $\beta$ band to
ground band which are typically overpredicted by an order of
magnitude.  The two exceptions where ES-D does not provide a good
description of energy spectra are $^{152}$Sm and $^{154}$Gd, which
have previously been shown \cite{CZ2,Tonev} to be well reproduced with the infinite
square well potential of the critical point symmetry X(5).
Furthermore, the ES-D model provides insights regarding the
recently found correlation \cite{Hinke} between the $\gamma$ stiffness and the
$\gamma$-bandhead energy. 

Concerning the long standing problem of
producing a level scheme with IBA SU(3) degeneracies within the
framework of the Bohr Hamiltonian, in the framework of the ES-D model
the spacings within the ground and $\beta$
bands are identical, because of the oscillator term in the
$u(\beta)$ potential. It is
therefore enough to examine the conditions under which the $2^+$,
$4^+$, $6^+$, \dots levels of the $\beta$ and $\gamma$ bands are
degenerate. One is then led \cite{ESD} 
to minimize the rms deviation between the even levels 
of the $\beta_1$ and $\gamma_1$ bands for fixed value of 
the normalized $\beta_1$-bandhead, $R_{0/2}= E(0_\beta^+)/E(2_1^+)$. 
Numerical results indicate that a
reasonable degree of degeneracy is obtained for $L_{max} =10$ and
$R_{0/2}\geq 15$, which is of physical interest, since experimental $R_{0/2}$
values extend up to 27~. 
In the case of $^{232}$Th, in which 
$R_{0/2}=14.8$, one can see that the above mentioned 
experimental rms deviation 
between the even levels of the $\beta_1$ and $\gamma_1$ bands is 
$\sigma_{\beta,\gamma}^{th}(L_{\max}=10)=1.142$, while
the corresponding theoretical quantity is 
$\sigma_{\beta,\gamma}^{exp}(L_{\max}=10)=0.593$ \cite{ESD}. Therefore,
although the overall fit is quite good, the degree of degeneracy
obtained from theory is less than the one indicated by experiment.
One could conclude that the ES-D model does contain parameter
pairs which correspond to approximate degeneracy of the low lying
even levels of the $\beta_1$ and $\gamma_1$ bands, while at the same
time the spacings within the $\beta_1$ band are identical to the
spacings within the ground band, but the problem of reproducing a
SU(3) spectrum from the Bohr Hamiltonian remains conceptually
open.

\section{Conclusion} 

The field of shape/phase transitions and critical point symmetries is 
rapidly expanding. Many additional references can be found in the review 
articles \cite{CastenNP,Poenaru,Rowe759}, as well as in Ref. \cite{LoBianco}.

\end{document}